\documentstyle[12pt]{article}

\def\DRAFT{0}

\newcommand{\eq}{\begin{equation}}
\newcommand{\en}{\end{equation}}

\newcommand{\eqa}{\begin{eqnarray}}
\newcommand{\ena}{\end{eqnarray}}
\newcommand{\eqan}{\begin{eqnarray*}}
\newcommand{\enan}{\end{eqnarray*}}

\newcommand{\lbl}[1]{ \ifnum \DRAFT = 0
                             {\label {#1}}
                      \else  {\mbox{\raisebox{-2ex}{\tiny #1}\hspace{-6ex}}}
                             {\label {#1}}
                      \fi}
\newcommand{\rf}[1]{  \ifnum \DRAFT = 0
                             \ref {#1}
                      \else {\ref {#1}}
                            {\raisebox{-2ex}{\tiny #1}}
                      \fi}

\newcommand{\sect}[1]{\setcounter{equation}{0}\section{#1}}

\newcommand{\draft}{ \ifnum \DRAFT =0  {} \else {\\ DRAFT} \fi}

\newcommand{\IJMP}[1]{Int. Jou. Mod. Phys. B\ {\bf #1}\ }

\newcommand{\MPL}[1]{Mod. Phys. Lett.\ {\bf #1}\ }

\newcommand{\NP}[1]{Nucl. Phys.\ {\bf #1}\ }

\newcommand{\PL}[1]{Phys. Lett.\ {\bf #1}\ }
\newcommand{\PR}[1]{Phys. Rev\ {\bf #1}\ }

\newcommand{\RMP}[1]{Rev. Mod. Phys.\ {\bf #1}\ }

\def\sqr#1#2{{\vcenter{\hrule height.#2pt
     \hbox{\vrule width.#2pt height#1pt \kern#1pt
        \vrule width.#2pt}
     \hrule height.#2pt}}}

\def\thinspace{\kern .16667em}

\def\Dir{\nabla\kern-2ex\Big{/}}
\def\dslash{\partial\kern-1.5ex\Big{/}}

\def\reali{{\hbox{\s@ l\kern-.5ex R}}}
\def\naturali{{\hbox{\s@ l\kern-.5ex N}}}
\def\interi{{\mathchoice
 {\hbox{Z\kern-1.5mm Z}}
 {\hbox{Z\kern-1.5mm Z}}
 {\hbox{{Z\kern-1.2mm Z}}}
 {\hbox{{Z\kern-1.2mm Z}}}  }}

\def\unity{{\hbox{\s@ 1\kern-.8mm l}}}
\def\uno{{\hbox{ 1\kern-.8mm l}}}
\def\part{\partial}

\def\rd{\sqrt{2}}
\def\um{{1\over2}}
\def\usrd{{1\over\sqrt{2}}}
\def\dxy{\delta^2(x-y)}
\def\dij{\delta^{ij}}

\def\lrarr{\leftrightarrow}
\def\rarr{\rightarrow}

\def\ot{\otimes}

\def\inf{\infty}

\def\CG{{\cal G}}

\def\CP{{\cal P}}

\def\CT{{\cal T}}

\def\aa{\alpha}
\def\bb{\beta}
\def\cc{\chi}

\def\dd{\delta}

\def\ee{\epsilon}

\def\ff{\phi}

\def\gg{\gamma}
\def\GG{\Gamma}

\def\pp{\psi}

\def\pb{\bar\psi}

\def\rr{\rho}

\def\ss{\sigma}
\def\SS{\Sigma}

\begin{document}

\begin{titlepage}

\begin{flushright}
NORDITA-94-35 P\\
July 1994\\
hep-th/9408018
\draft
\end{flushright}
\vspace*{0.5cm}

\begin{center}
{\bf
\begin{Large}
{\bf
GENERALIZED QCD$_2$ VIA THE BILOCAL METHOD
\\}
\end{Large}
}
\vspace*{1.5cm}
          {\large Igor Pesando}
           \footnote{E-mail PESANDO@NBIVAX.NBI.DK, 22105::PESANDO,
                     31890::I\_PESANDO.}
           \footnote{Work supported by the EU grant ERB4001GT930141.}
         \\[.3cm]
          NORDITA\\
          Blegdamsvej 17, DK-2100 Copenhagen \O \\
          Denmark\\
\end{center}
\vspace*{0.7cm}
\begin{abstract}
{
We use the bilocal method to derive the large $N$ solution of the most
general $QCD_2$.
}
\end{abstract}
\vfill
\end{titlepage}

\setcounter{footnote}{0}
\newcommand{\g}[4]{ {\cal G}_{ {{#1} {#2}},{{#3} {#4}} } }
\def\CGt{{\cal G}^{tt}}
\newcommand{\CGT}{ {\mathop{\cal G}\limits _t} }
\newcommand{\CGS}{ {\mathop{\cal G}\limits _s} }

\newcommand{\jj}[2]{ J_{{#1} {#2}} }

\newcommand{\CMJ}{ {\mathop{\cal M}\limits _J} }

\newcommand{\U}[7]{ {\mathop{U}\limits _{#1}}{}_{{#2} {#3};{#4}{#5},{#6}{#7}} }
\newcommand{\UZ}[2]{ {\mathop{\cal U}\limits_{0}} {}_{{#1} {#2}} }
\newcommand{\UU}[4]{ {\mathop{\cal U}\limits_{1}} {}_{{#1} {#2};{#3}{#4}} }
\newcommand{\UD}[6]{ {\mathop{\cal U}\limits_{2}}
                     {}_{{#1}{#2};{#3}{#4},{#5}{#6}} }
\newcommand{\MMZ}{ {\mathop{\cal M}\limits_{0}} }
\newcommand{\UUZ}{ {\mathop{\cal U}\limits_{0}} }
\newcommand{\UUU}{ {\mathop{\cal U}\limits_{1}}}
\newcommand{\UUD}{ {\mathop{\cal U}\limits {}_{2}}}

\newcommand{\SZ}[2]{ {\mathop{\Sigma}\limits_{0}} {}_{{#1} {#2}}^{(1)} }
\newcommand{\SU}[4]{ {\mathop{\Sigma}\limits_{1}}
                     {}_{{#1} {#2};{#3}{#4}}^{(1)} }
\newcommand{\SSZ}{ {\mathop{\Sigma}\limits_{0}}{}^{(1)} }
\newcommand{\SSU}{ {\mathop{\Sigma}\limits_{1}}{}^{(1)} }

\newcommand{\GU}[2]{ {\mathop{G}\limits_{1}} {}_{{#1} {#2}} }
\newcommand{\GD}[4]{ {\mathop{G}\limits_{2}} {}_{{#1} {#2};{#3}{#4}} }
\newcommand{\GSU}{ {\mathop{G}\limits_{1}} }
\newcommand{\GSD}{ {\mathop{G}\limits_{2}} }

\newcommand{\uu}[2]{ U_{{#1} {#2}} }
\newcommand{\uj}[2]{ {\mathop{U}\limits _J}{}_{{#1} {#2}} }
\newcommand{\uuj}{ {\mathop{U}\limits _J} }
\newcommand{\du}[2]{\delta U_{{#1} {#2}} }

\newcommand{\mm}[2]{ M_{{#1} {#2}} }
\newcommand{\mj}[2]{ {\mathop{M}\limits _J}{}_{{#1} {#2}} }
\newcommand{\mmj}{ {\mathop{M}\limits _J} }
\newcommand{\dm}[2]{\delta M_{{#1} {#2}} }
\def\DT{{\tilde D}}
\def\UT{{\tilde U}}
\def\PIT{{\tilde \Pi}}
\def\SST{{\tilde \Sigma}}
\def\zt{{\tilde z}}
\def\dz{{\sqrt{2}z}}

\def\uij{U_{ij}}
\def\ucij{U^\dagger_{ij}}
\def\uji{U_{ji}}
\def\ucji{U^\dagger_{ji}}
\def\zpm{z_{\pm}}
\def\zp{z_+}
\def\zm{z_-}
\def\kp{k_+}
\def\km{k_-}
\def\lp{l_+}
\def\lm{l_-}
\def\ddt{{\dd T}}
\def\mucr{\mu_{cr}}

\newcommand{\mat}[4]{\left(
                     \begin{array}{cc}
                     {#1} & {#2} \\
                     {#3} & {#4}
                     \end{array}
                     \right)
                    }
\newcommand{\vett}[2]{\left(
                      \begin{array}{c}
                     {#1} \\
                     {#2}
                     \end{array}
                     \right)
                    }
\newcommand{\vet}[2]{\left(
                     \begin{array}{cc}
                     {#1} &  {#2}
                     \end{array}
                     \right)
                    }
\newcommand{\ft}[3]{\int {d^{#1}{#2}\over (2\pi)^{#1}} ~ e^{i {#2}\cdot{#3}} }

\def\rhm{\rho_-}
\def\rhp{\rho_+}
\def\sgm{\sigma_-}
\def\sgp{\sigma_+}

\def\rd{\sqrt{2}}
\def\usrd{{1\over\sqrt{2}}}
\def\dxy{\delta^2(x-y)}
\def\dij{\delta^{ij}}
\def\dsi{\partial_{x^-}}
\def\dta{\partial_{x^+}}

\newcommand\modu[1]{|{#1}|}

\newcommand\psiind[3]{\psi^{#1 #2}_{#3}}
\newcommand\psibind[3]{{\bar\psi}^{#1 #2}_{#3}}
\newcommand\psiindd[2]{\psi^{#1 #2}}
\newcommand\psibindd[2]{{\bar\psi}^{#1 #2}}

\def\pbai{\psibindd{A}{i}}
\def\pbaj{\psibindd{A}{j}}
\def\pbbi{\psibindd{B}{i}}
\def\pbbj{\psibindd{B}{j}}
\def\pai{\psiindd{A}{i}}
\def\paj{\psiindd{A}{j}}
\def\pbi{\psiindd{B}{i}}
\def\pbj{\psiindd{B}{j}}

\def\paip{\psiind{A}{i}{+}}
\def\pajp{\psiind{A}{j}{+}}
\def\pbip{\psiind{B}{i}{+}}
\def\pbjp{\psiind{B}{j}{+}}
\def\paim{\psiind{A}{i}{-}}
\def\pajm{\psiind{A}{j}{-}}
\def\pbim{\psiind{B}{i}{-}}
\def\pbjm{\psiind{B}{j}{-}}

\def\pbaip{\psibind{A}{i}{+}}
\def\pbajp{\psibind{A}{j}{+}}
\def\pbbip{\psibind{B}{i}{+}}
\def\pbbjp{\psibind{B}{j}{+}}
\def\pbaim{\psibind{A}{i}{-}}
\def\pbajm{\psibind{A}{j}{-}}
\def\pbbim{\psibind{B}{i}{-}}
\def\pbbjm{\psibind{B}{j}{-}}

\newcommand\fai[4]{{#1}^{{#2}}_{{#3}} ({#4}) }

\def\rjimyx{\fai{\rho}{j i}{-}{y,x}}
\def\rijmxy{\fai{\rho}{a b}{-}{x,y}}
\def\rikmxx{\fai{\rho}{i k}{-}{x,x}}
\def\rikmxz{\fai{\rho}{i k}{-}{x,z}}
\def\rkjmxy{\fai{\rho}{k j}{-}{x,y}}
\def\rkjmzy{\fai{\rho}{k j}{-}{z,y}}
\def\rljmxy{\fai{\rho}{k j}{-}{x,y}}
\def\rllmuu{\fai{\rho}{l l}{-}{u,u}}
\def\rijpxy{\fai{\rho}{a b}{+}{x,y}}
\def\rkjpxy{\fai{\rho}{k j}{+}{x,y}}
\def\rkjpzy{\fai{\rho}{k j}{+}{z,y}}
\def\rijpxy{\fai{\rho}{i j}{+}{x,y}}
\def\rikpxy{\fai{\rho}{i k}{+}{x,y}}
\def\rikpxz{\fai{\rho}{i k}{+}{x,z}}
\def\rikpyz{\fai{\rho}{i k}{+}{y,z}}
\def\rikpxx{\fai{\rho}{i k}{+}{x,x}}

\def\sijlxy{\fai{\ss}{a b}{L}{x,y}}
\def\siklxx{\fai{\ss}{i j}{L}{x,x}}
\def\siklxz{\fai{\ss}{i j}{L}{x,z}}
\def\skjlxy{\fai{\ss}{k j}{L}{x,y}}
\def\skjlzy{\fai{\ss}{k j}{L}{z,y}}

\def\sijrxy{\fai{\ss}{a b}{R}{x,y}}
\def\sikrxx{\fai{\ss}{i j}{R}{x,x}}
\def\sikrxz{\fai{\ss}{i j}{R}{x,z}}
\def\skjrxy{\fai{\ss}{k j}{R}{x,y}}
\def\skjrzy{\fai{\ss}{k j}{R}{z,y}}

\def\aijp{{\alpha^{i j}_+}}
\def\aijm{{\alpha^{i j}_-}}
\def\bijp{{\beta ^{i j}_L}}
\def\bijm{{\beta ^{i j}_R}}

\def\aijmxy{\fai{\aa}{i j}{-}{x,y}}
\def\aijpxy{\fai{\aa}{i j}{+}{x,y}}
\def\ajimyx{\fai{\aa}{j i}{-}{y,x}}
\def\ajipyx{\fai{\aa}{j i}{+}{y,x}}

\def\bijpxy{\fai{\bb}{i j}{L}{x,y}}
\def\bijmxy{\fai{\bb}{i j}{R}{x,y}}
\def\bjipyx{\fai{\bb}{j i}{L}{y,x}}
\def\bjimyx{\fai{\bb}{j i}{R}{y,x}}
\def \bigd{{\cal D}}

\def\ab{{\bar a}}

\sect{Introduction}

One of the most interesting and still not fully understood problem of
Quantum Field Theory concerns the calculation of the mass of relativistic
bound states. This problem is even more complicated to solve in the case of
$QCD$, the $SU(3)$ gauge theory of quarks and gluons, that describes the
strong interaction physics where the basic constituents are confined and
perturbation theory cannot be used. The expectation is that  the
low energy spectrum consists of colorless mesons and baryons, but up to
now the only available method seems to be based on numerical calculation
using the lattice theory formulation of the theory.

The large-$N$ expansion technique proposed by 't Hooft~\cite{tH1,tH}
several years ago seems to be the most promising approach to obtain
analytic results on the hadron spectrum.
It is well known that in the limit
in which the number $N$ of colours becomes very large the theory becomes
much simpler, in the sense that only planar Feynman graphs survive.
Furthermore, when $N \rightarrow \infty$ the theory only contains colorless,
stable and noninteracting mesons with two-body decay and scattering amplitudes
proportional to ${1\over \sqrt{N}}$ and ${1\over N}$, respectively (there
is also a way of describing baryons as solitons of the effective Lagrangian
in the large $N$ limit, but this will not interest us in this context).

Unfortunately, all efforts to try to solve four-dimensional large-$N$
$QCD$ have failed up to now.
The main problem is to find the semi-classical configuration,
namely the {\it master field}~\cite{Witten}, from which the action is
dominated in the large-$N$ limit and whose fluctuations around the vacuum
should give the particle spectrum of the theory.

The main reason of this failure is due to the fact that no method
has yet been found to solve matrix models for a space time dimension $D
> 2$.
On the other side it is well known that the large $N$ expansion
can be explicitly
performed in vector models as for instance the $O(N)$ vector model (see
for instance ref.~\cite{PDV} and references therein),
the two-dimensional $CP^{N-1}$ model~\cite{CPN}
and also in $QCD_2$~\cite{tH,CCG} and \cite{Gut,CDP,indiani} with matter in
the fundamental representation of the gauge group
Some results can be obtained with both fermionic and bosonic adjoint matter
\cite{KUT} though the model is a real 2D matrix model.

In particular this set of models are solved by means of two slightly
different methods. The vector-like models are solved in the large $N$
expansion by introducing
a local "colourless" composite field and by explicitly integrating over the
fundamental fields (See for instance Refs.\cite{PDV} and \cite{CPN}):
with the help of the composite local field we can extract the explicit N
dependence and apply the saddle point technique.

The $QCD_2$-like models are characterized by the fact that the number of
components of the non abelian gauge field goes to infinity in the large
N expansion and from this viewpoint they are matrix models.
The gauge field in two dimensions has no physical degrees
of freedom and therefore can be eliminated by using its classical
equation of motion. In this way one gets a non local Coulomb interaction
that is quartic in terms of the vector-like matter fields. The theory
can then be solved in the large N limit  by introducing a bilocal composite
field as discussed in Refs.~\cite{Gut,CDP,indiani}.
We can rephrase this by saying that the kernel of the solubility lies in the
fact that the matrices of the theory (the gauge fields) are expressible as
tensor product of vectors.
The same happens in the generalized QCD \cite{DLS}, even if the theory seems a
2
matrices model, it is actually a nonlocal vector model.


In this paper we present a solution of the most general case of generalized
$QCD_2$ using the bilocal method (see \cite{CAV} for an extensive application
of
the method to the other known cases).

\sect{The master field of the general QCD$_2$:the simplest case.}

The usual $QCD_2$ action can be written in the first order formalism as
\footnote{{\bf Conventions.}
$$
x^\pm=x_\mp=\usrd(x^0\pm x^1)~~~~
A^\mu B_\mu=
A_0 B_0 - A_1 B_1=
A_+ B_- + A_+ B_-
{}~~~~
\ee^{01}=-\ee^{+-}=1
$$
$$
\gg_+=\mat{0}{\rd}{0}{0}~~~~
\gg_-=\mat{0}{0}{\rd}{0}~~~~
\gg_0=\mat{0}{1}{1}{0}~~~~
\gg_1=\mat{0}{1}{-1}{0}~~~~
$$
$$
\gg_5=-\gg_0\gg_1=\mat{1}{0}{0}{-1}~~~~
P_{R,L}={1\pm\gg_5\over2}
$$
$$
\pp=\vett{\pp_+}{\pp_-}~~~~
\pb=\vet{\pb_-}{\pb_+}~~~~
 \cc\pb=-\usrd
\mat
{\rd\pb P_R\cc}
{\pb \gg_-\cc}
{\pb \gg_+\cc}
{\rd\pb P_L\cc}
$$
$$
\int_x=\int d^2 x ~~~~
\int_p=\int {d^2 p\over (2\pi)^2}
$$
}

\eq
S=\int d^2x~
\left\{
{N\over 8\pi}tr_c(E{\tilde F})
-{N g^2\over 4\pi}tr_c(E^2)
+\pb^a(i\Dir-m^i\uno)\pp^a
\right\}
\lbl{qcd2}
\en
where $F_{\mu\nu}=\part_\mu A_\nu-\part_\nu A_\mu +i[A_\mu,A_\nu]$,
${\tilde F}=F_{\mu\nu}\ee^{\mu\nu}$,
$i\Dir_{A B}=\gg_\mu(i\part_\mu \uno_{A B}-A^\ab_\mu T^\ab_{A B})$,
$\ab,{\bar b}=1...N^2$ are the indices of the adjoint representation of the
colour group, $A,B=1...N $ run over the fermionic representation
of the colour $U(N)$ and $a,b...$ are flavour indices, and $tr_c$ is
the color trace.

{}From the previous form it can be directly generalized to (\cite{DLS})
\eq
S=\int d^2x~
\left\{
{N\over 8\pi}tr_c(E{\tilde F})
-{N g^2\over 4\pi}\sum_n ~f_n tr_c(E^n)
+\pb^a(i\Dir-m^i\uno)\pp^a
\right\}
\lbl{s-prima}
\en

To solve this theory we choose the usual gauge $A^\ab_-=0$.
and we can rewrite the previous expression as
\eqa
S=\int&d^2x&
\biggl\{
{N\over 4\pi}~tr_c(E \part_- A_+)
-{N g^2\over4\pi}\sum_{n=2}^{\inf} f_n tr_c(E^n)
\nonumber\\
&+&
\pb^a(i\dslash-m_i)\pp^a
-A_+^\ab (\pb^a\gg_- T^\ab\pp^a)
\biggr\}
{}~\lbl{s-gauge}
\ena
where $tr_a$ and $tr_\aa$ are the traces over the flavour and spin indices.

If we integrate over $A^\ab_+$, we get the constraint
\eq
\dsi\left({N\over 4\pi}~E^\ab\right)
+\pb^a\gg_- T^\ab\pp^a=0
\lbl{const0}
\en
in the form of a delta function in the path integral.
Since $E$ is in the adjoint representation of $U(N)$, the constraint
(\rf{const0}) can be rewritten as
\eq
\dsi\left({N\over 4\pi}~E^{A B}\right)
+\pb^{B a}\gg_- \pp^{A a}=0
\lbl{const}
\en
Now we can integrate easily over $E^{A B}$ getting the effective action
for the $\pp$ fields
\eqa
S_{eff}&=&
\int d^2x~\biggl\{
 \pb^{A a}(i\dslash-m_a)\pp^{A a}
\nonumber\\
&-&{N g^2\over 4\pi}\sum_n ({4\pi\over N})^n f_n
(\part_-^{-1}\pb^{A_2 a_1}\gg_-\pp^{A_1 a_1})_x
\dots
(\part_-^{-1}\pb^{A_1 a_n}\gg_-\pp^{A_n a_n})_x
\biggr\}
\nonumber\\
{}~
\ena

The interaction term suggests to introduce the (composite) bilocal field
\eq
U_{a\aa,b\bb}(x,y)=U_{P Q}=-{\rd\over N}tr_c(\pp^a_\aa(x)~\pb^b_\bb(y))
=\mat{\sijrxy}{\rijmxy}{\rijpxy}{\sijlxy}
\lbl{u}
\en
where $P=(x a \aa), Q=(y b \bb)$, and
\eq
P_{P Q}(u)=P_{x a\aa,y b\bb}(u)
=\left(-{4\pi\over\rd}\right) \dxy\part^{-1}_-(u,y)~\dd^{a b}~(\gg_-)_{\aa\bb}
\en
The effective action can then be rewritten as
\eqa
{1\over N}S_{eff}
&=&
\int_{x,y}
\usrd\dxy
tr_{a,\aa}\left( (i\dslash_x-m_a)U(x,y)\right)
+{g^2\over 4\pi}\sum_n f_n\int_{x}
Tr\left(P(x)U\right)^n
\nonumber\\
&=&Tr\left(D~U+{g^2\over4\pi}\sum_n f_n\int_x (P(x)U)^n~\right)
\ena
where $Tr=tr_P=tr_x~tr_a~tr_\aa$,
and
\eqa
D=||D_{P Q}||&=&
\mat
{-\usrd m_a~\dd^{a b}~\dxy}
{i~\dd^{a b}~\part_{x^-}\dxy}
{i~\dd^{a b}~\part_{x^+}\dxy}
{-\usrd m_a\dd^{a b}\dxy}
\nonumber\\
{}~
\ena

In order to get the effective action for the field $U$, we have to
compute the jacobian of the transformation $J[U]$ from $\pp$ to $U$.
This can be accomplished as in Ref.s (\cite{CDP,DAS}).
The result is
\eq
J[U]
\propto\int [d M]
\exp\left\{N [i~Tr(MU)+Tr \log M]\right\}
\en

The final form for the effective action is given as a functional of the $U$
and $M$ fields, and it is
\eq
{1\over N}S_{eff}
=Tr\left(D~U+{g^2\over4\pi}\sum_n f_n\int_x [P(x)U]^n
+ M~U-i \log M\right)
\lbl{s-eff}
\en

{}From this action we can immediately compute the saddle point equations:
\eq
\left\{
\begin{array}{l}
\UUZ~\MMZ-i\uno=0
\\
D+{g^2\over4\pi}\sum_n n f_n\int_x [P(x)\UUZ]^{n-1}P(x)+\MMZ=0
\end{array}
\right.
\en
that yield
\eq
D~\UUZ+{g^2\over4\pi}\sum_n n f_n\int_x (P(x)\UUZ)^{n}
=-i\uno
\lbl{sp}
\en
Following the approach used in the usual case (\cite{CDP}), we introduce
the self-energy $\SS=\um \mat{\GG_L}{\GG_-}{\GG_+}{\GG_R}$ through the
definition
\eq
\UUZ=-i(D+\SS)^{-1}
\en
Inserting it in eq. (\rf{sp}) we get the following eq. for $\SS$
\eq
\SS= {g^2\over4\pi}\sum_n n f_n\int_x
\left(P(x){-i\over D+\SS}\right)^{n-1}P(x)
\lbl{sp-sigma}
\en
that yields immediately $\GG_L=\GG_-=\GG_R=0$ because of the $\gg_-$
included in $P(x)$.
Assuming translational invariance for the ground state as in Ref.
(\cite{CDP}), we can write the
master field as
\eqa
\UZ P Q
&=&\int_{p,q}e^{ipx+iqy}~\UUZ^{a b}_{\aa\bb}(p,q)
\nonumber\\
\UUZ^{a b}_{\aa\bb}(p,q)&=&(2\pi)^2\dd(p+q)
 {i~\dd^{a b}\over p^2-m^2_a-p_-\GG_{a +}(p)+i\ee}
\mat{-\rd m_a}{2 p_-}{2 p_+ -\GG_{a +}(p)}{-\rd m_a}_{\aa\bb}
\nonumber\\
{}~\lbl{u0}
\ena
Using the previous eq. (\rf{u0}) the eq. (\rf{sp-sigma}) can be easily
rewritten in the momentum space as
\eq
\GG^{a }_+(p)= 2g^2 \sum_n (-i)^n n f_n I_n^{a}(p,p)
\lbl{gg}
\en
where
\eq
\dd_{a b}I^a_n(p,q)= (4\pi)^{n-1}
\int_{p_1\dots p_{n-1}}
{1\over (p -p_1)_-}
\rr^{a k_1}_-(p_1)
{1\over (p_1 -p_2)_-}\dots
\rr^{k_{n-1} b}_-(p_{n-1})
{1\over (p_{n-1}-q)_-}
\en
where $\rr_-$ is defined in eq. (\rf{u}).

Using the principal value regularization for $1\over k_-$ as in
Ref. (\cite{DLS}) and symmetric $p_+$ integration done by 't Hooft
(\cite{tH}, we get the result
\eqa
I_n^{a }(p,p')&=&\int^\infty_{-\infty} dk_1 \dots dk_{n-1}
\nonumber\\
&&
{\CP \over p-k_1} sgn(k_1)
{\CP \over k_1-k_2} sgn(k_2)
\dots
{\CP \over k_{n-2}-k_{n-1}} sgn(k_{n-1})
{\CP \over k_{n-1}-p'}
\nonumber\\
{}~
\ena
Notice the useful property: $I_n(\aa p,\aa p')={1\over \aa} I_n(p,p')$, that
will turn out to be useful while computing the full $q{\bar q}$ amplitude in
the
next section.

\sect{The full quark-antiquark amplitude}

In order to compute the full $q {\bar q}$ amplitude we add a source $J$
coupled to the composite field $U$ to the action.
The partition function in presence of an external source $J$ is given
by:
\eq
Z[J]=Z_J=
\int [d U][d M] e^{iNS_J}
\en
where
\eq
S_J=S_{eff}-i~Tr(J U)
\en
Differentiating wrt $J$ the logarithm of the partition function, we find
the connected Green functions (from now on all the capital roman indices
will be used to indicate collection of indices $(x a \aa)$)
\eqa
\GU A B=
<\uu A B>_c={1\over N}\left.{ \dd \log Z_J\over \dd \jj B A}\right|_{J=0}
\\
\GD A B C D=
<\uu A B\uu C D>_c={1\over N^2}{ \dd^2 \log Z_J\over \dd\jj B A \dd\jj D C}
\ena
To the leading order in $N$ we have obviously that
\eqa
\log(Z_J^{(0)})=i N S_J[\uj{}{}]
\\
<\uu A B>_c^{(0)}=\UZ A B
\\
<\uu A B\uu C D>_c^{(0)}={1\over N}\UU A B C D
\lbl{def-u1}
\ena
where $\uj{}{}$ is the saddle point value in presence of the source,i.e.
\eq
\uj A B =\UZ A B+\UU A B P Q \jj Q P
+\dots
\lbl{uj-exp}
\en
and it is given by the solution  of the equation generalizing (\rf{sp}) in the
presence of $J$
\eq
D +i\uuj^{-1}
+{g^2\over4\pi}\sum_n n f_n\int_x (P(x)\uuj)^{n-1} P(x)
+J=0
\lbl{sp-j}
\en
This equation cannot be solved exactly for an arbitrary $\jj{}{}$, but it can
be solved perturbatively in $\jj{}{}$.
Inserting  (\rf{uj-exp}) in the previous equation and using
\eq
\uj A B ^{-1}
\approx
\biggl(
\UZ{}{}^{-1}
-\UZ{}{}^{-1}\UU{}{}P Q\UZ{}{}^{-1} \jj Q P
\biggr)_{ A B}
\en
we find
\eqa
&D&+ i~\UUZ ^{-1}
+{g^2\over4\pi}\sum_n n f_n\int_x (P(x)\UUZ)^{n-1}P(x)
=0
\nonumber\\
{}~
\ena
and
\eqa
&-&i\UUZ^{-1}\UU{}{}P Q\UUZ^{-1}
-iX_{;Q P}+
\nonumber\\
&&+{g^2\over4\pi}\sum_{n=2} n f_n
               \sum_{m=0}^{m=n-2}\int_x \left[(P(x)\UUZ)^{m} P(x)\right]
                                        \UU{}{}P Q
                                        \left[(P(x)\UUZ)^{n-m-2}P(x)\right]
=0
\nonumber\\
{}~
\ena
where $X_{A B;P Q}=\dd_{A Q} \dd_{B P}$
The first of these equation has as solution the expression in eq. (\rf{u0}),
while the other can be formally solved as follows\footnote{
We use the convention that
all the matrix operations written in capital letter refer to the mesonic
indices, i.e. couples $(AB)\dots$, while the others to the quark ones, i.e.
$A,B\dots$.
Explicitly $(A^T)_{A B,C D}=A_{C D,A B}$
is the transposition wrt to the mesonic indices
while $(X^t)_{A B}=X_{B A}$ is wrt to the quark ones.

We use also $(X\ot Y)_{A B,C D}=X_{A C} Y_{B D}$ and
$(A_{; P Q}\ot Y)_{A B,C D}=A_{A C; P Q} Y_{B D}$ and
$(A_{;P Q}X)_{A B}=A_{A C;P Q} X_{C B}$.
}
\eq
\hspace*{-5em}
\UUU
=-\left(\uno
+i{g^2\over4\pi}\sum_{n=2} n f_n
               \sum_{m=0}^{m=n-2}\int_x \left[(P(x)\UUZ)^{m+1}\right]
                                        \ot
                                        \left[(P(x)\UUZ)^{n-m-1}\right]^t
\right)^{-1}\UUZ\ot\UUZ^t X
\lbl{u1}
\en

If we define
\eq
i \UUZ\ot\UUZ^t X\CG=
i{g^2\over4\pi}\sum_{n=2} n f_n
               \sum_{m=0}^{m=n-2}\int_x \left[(P(x)\UUZ)^{m+1} \right]
                                        \ot
                                        \left[(P(x)\UUZ)^{n-m-1}\right]^t
\en
the previous equation for $\UUU$ can be rewritten in two different revealing
forms as
\eqa
\UUU
&=&-(\UUZ\ot\UUZ^t)X
-(\UUZ\ot\UUZ^t X)~i\CG~ \UUU
\nonumber\\
&=&-(\UUZ\ot\UUZ^t)X
+i(\UUZ\ot\UUZ^t X)
  \left[\CG (\uno+i \UUZ\ot\UUZ^t X\CG)^{-1}\right]
  \UUZ\ot\UUZ^t X
\lbl{u1-ric}
\ena
The first recursive form reveals clearly that the role of the usual gluonic
interaction is now played by $\CG$.
Moreover the second form shows that
\eq
\CT=\CG(\uno+iX\UUZ^t\ot\UUZ\CG)^{-1}
\lbl{t}
\en
is the full quark-antiquark truncated amplitude and it is the generalization
of the full $q{\bar q}$ amplitude of Callan, Curtis and Gross $\CT_{CCG}$
(\cite{CCG}), as it is transparent from the following recursion
relation, that it satisfies
\eq
\CT=
\CG
-i\CG X (\UUZ^t\ot\UUZ)\CT
\lbl{rec-rel}
\en
Notice however that the recursion relation for $\CT$ has a minus sign
in front of the second term on rhs of difference wrt
the CCG one: this has no consequences and it is due to the fact that we
are dealing with an effective theory where the gluonic degrees of
freedom are integrated away. This implies the appearance of quartic fermionic
vertices and hence the substitution of every loop quark-gluon-quark-gluon in
the original theory by a fermionic loop due to the
elimination of the gluons. Since every fermionic loop carries a minus
sign, this explains the origin of the difference.

Now we go to the momentum space and we write
the "gluonic" interaction $\CG$ and the full $q{\bar q}$ amplitude
 explicitly as
\eqa
\CG_{A B;C D}
\equiv
\CG^{a b, c d}_{\aa\bb,\gg\dd}[p,q,q',p']
=(2\pi)^2\dd^2(p+p'+q+q')~\dd^{b c}\dd^{d a}~(\gg_-)_{\bb\gg}(\gg_-)_{\dd\aa}
 G^{a b}(p,p';r=p+q)
\nonumber\\
{}~
\ena
\eq
 G^{a b}[p,q,q',p']=
2\pi g^2~\sum_{n=2} (-i)^n n f_n
                       \sum_{m=1}^{m=n-1} I^b_{m}(q,-q') I^a_{n-m-1}(p',-p)
\en
and
\eqa
\CT_{A B;C D}
\equiv
\CT^{a b, c d}_{\aa\bb,\gg\dd}[p,q,q',p']
=(2\pi)^2\dd^2(p+p'+q+q')~\dd^{b c}\dd^{d a}~(\gg_-)_{\bb\gg}(\gg_-)_{\dd\aa}
 T^{a b}(p,p';r)
\nonumber\\
{}~
\ena
where all the momenta are incoming.
The recursion relation (\rf{rec-rel})  reads now as
\eqa
 T^{a b}(p,p';r)
&=&G^{a b}(p,p';r)
\nonumber\\
&-&2i\int_m G^{a b}(p,-m;r) \rr^a_-(-m) \rr^b_-(r-m)
          T^{a b}(m,p';r)
\lbl{rec-rel-expl}
\ena
Defining as usual the $x$ and $x'$ variables as $p_-=x r_-$, $p_-'=-x' r_-$
(the minus is due to the fact
that our momenta are all incoming and we want to make contact with the usual
conventions) and
\eq
\ff(x,x';r)=
{i\over2}\int d p_+~\rr^a_-(-p)\rr^b(r-p) T^{a b}(p,p';r)
\en
the previous equation (\ref{rec-rel-expl}) yields $T$ as a functional of $\ff$
\eq
T^{a b}(p,p';r)
={1\over r_-^2}G^{a b}(x,-x';1)
-{1\over \pi^2 r_-}\int d y~G^{a b}(x,y;1)\ff^{a b}(y,-x';r)
\en
Defining the eigenfunctions $\ff^{a b}_n(x)$ through the equation
\eqa
M^2_{(a b)n} \ff^{a b}_n(x,x';r)
&=&
\left[ {m_a^2+x \GG_+(x)\over x}+ {m_b^2+(1-x)\GG_+(1-x)\over 1-x}\right]
 \ff^{a b}_n(x,x';r)
\nonumber\\
&&-\int^1_0 dy~{2\over \pi} G^{a b}(x,-y;1)\ff^{a b}_n(y,x';r)
\ena
and supposing that they form a complete system in $x\in[0,1]$, we can write
the full amplitude $T$ as
\eqa
\begin{array}{c}
{
T^{a b}(p_-,p_-';r)={1\over r_-^2}G^{a b}(x ,-x';1)
}

\\
{
-{4\over \pi^2 r_-^2}\sum_n{1\over r^2-M_{(a b)n}^2}
\int^1_0dy~G^{a b}(x,-y;1) \ff^{a b}_n(y;r)
\int^1_0dy'~G^{a b}(y',-x';1) \ff^{a b *}_n(y';r)
}
\end{array}
\nonumber\\
{}~
\ena
Specializing the previous equation to the normal case, we
notice that $T_{CCG}=2i~T$; the factor $i$ can be traced back to the
definition (\rf{t}) and the factor $2$ is due to the definition of $U$
(\rf{u}) (see also (\rf{def-u1})).

\sect{The general case.}
The most general expression for the generalized $QCD_2$ is given by the action
\eq
S=\int d^2x~
\left\{
{N\over 8\pi}tr_c(E{\tilde F})
-{N g^2\over 4\pi}\sum_{\{n_i\}} ~f_{\{n_i\}} \prod_{n\in\{n_i\}} tr_c(E^n)
+\pb^a(i\Dir-m^i\uno)\pp^a
\right\}
\lbl{s-gen}
\en
where ${\{n_i\}} $ is a whatever sequence of natural number satisfying the
condition $n_i\le n_{i+1}$.
Repeating the previous steps we find the effective action
\eq
{1\over N}S_{eff}
=Tr\left(D~U+{g^2\over4\pi}\sum_{\{n_i\}}
 f_{\{n_i\}}\int_x \bigl[\prod_{n\in\{n_i\}}(P(x)U)^n\bigr]
+ M~U-i \log M\right)
\lbl{s-gen-eff}
\en
and the saddle point condition
\eq
D~\UUZ+{g^2\over4\pi}\sum_{\{n_i\}} f_{\{n_i\}}
\sum_{n\in\{n_i\}}\int_x (P(x)\UUZ)^{n}\prod_{m\in\{n_i\} \setminus \{n\}}
                        Tr(P(x)\UUZ)^{m}
=-i\uno
\lbl{sp-gen}
\en
Using the translational invariance of the saddle point, i.e. of the ground
state, it turns out that
\eq
T_n=Tr(P(x)\UUZ)^n=
4\pi\left(-{\rd\over4\pi i}\right)^n\int_p\sum_a I^a_n(p,p)\rr^{a a}_-(p)
\en
is independent of $x$, this means that the whole effect of introducing the most
general potential amounts in generating
the effective coupling constants ${\bar f}_n$ given by
\eq
{\bar f}_n
={1\over T_n}\sum_{\{n_i\}\mbox{ such that } n\in\{n_i\} }
f_{\{n_i\}}
d_n({\{n_i\}})
\prod_{m\in\{n_i\}} T_m
\lbl{f-eff}
\en
where $d_n({\{n_i\}})$ is the degeneration of $n$ in the sequence ${\{n_i\}}$.
As it can be shown easily all the formula of the previous section are still
valid with the substitution $f_n \rarr {\bar f}_n$ (\ref{f-eff}).

We notice however that the theory is independent of ${\bar f}_1$ because
$\int_x P(x)=0$ as a consequence of $\int_x P(x)\propto a_+\gg_-$
and of the absence of a constant vector $a_+$ in the theory.

Moreover we have
\eq
T_{2n+1}=0
\lbl{t-odd}
\en
and this implies that it is impossible to induce terms with odd powers starting
from terms with even total powers.
The demonstration of (\ref{t-odd}) goes as follows:
\eq
T_{n+1}\propto
\int_{p,p_1\dots p_n}
{1\over (p -p_1)_-}
\rr^{a c_1}_-(p_1)
{1\over (p_1 -p_2)_-}\dots
\rr^{c_n b}_-(p_{n-1})
{1\over (p_n-p)_-}
\rr^{b a}_-(p)
\en
then we rename the integration variables as $p\lrarr p_n, p_1\rarr p_{n-1},
p_2\rarr p_{n-2}\dots$ and we use $\rr_-(-p)=-\rr_-(p)$ (that follows from
(\ref{gg},\ref{u0})), and we find $T_n=(-)^n T_n$, that proves what asserted,
independently of the regularization scheme.

\sect{Conclusions.}

In this work we have solved the most general case of minimally coupled
generalized $QCD_2$ (\cite{DLS}) using the bilocal method (\cite{DAS,CDP}).
This is achieved in a straightforward and transparent way, thus showing the
effectiveness of the method, when it is applied to vector-like models
coupled to matrix gluon fields, which in two dimensions are not dynamical
(see also \cite{CAV}):
the main point that guarantees the successful application of the large N
techniques is the possibility of defining colourless fields
(see eq. (\ref{u}) ) that allows one to extract
the $N$ dependence both in the action and in the measure of integration
(the mesonic bilocal $U(x,y)$ is a global colour singlet,
as it is the auxiliary bilocal field $M(x,y)$).

\end{document}